\documentclass[preprint,showpacs,preprintnumbers,amsmath,amssymb]{revtex4}

\usepackage{graphicx}
\usepackage{dcolumn}
\usepackage{bm}

\begin{document}


\title{Is Kinetic Growth Walk equivalent to canonical Self Avoiding Walk?}

\author{M. Ponmurugan$^1$, S. L. Narasimhan$^{2*}$ and K. P. N. Murthy$^1$ \\ 
$^1$Materials Science Division, Indira Gandhi Center for Atomic Research, Kalpakkam - 603 102, Tamil Nadu, India. \\
$^2$Solid State Physics Division, Bhabha Atomic Research Center, Mumbai - 400 085, India.}


\begin{abstract}
We present a Monte Carlo study of Kinetic Growth Walk on square as well as triangular lattice to show that it is not equivalent to canonical Self Avoiding Walk. 
\end{abstract}

\maketitle

\section*{Introduction}
Growth is a natural process by which a system, for example, a tree acquires a history-dependent (or a {\it causal}) structure. Since a growing tree is never ungrown, its growth is an irreversible process. A generic but informal description of its structural features evolves from repeated observations of a variety of structural forms that these systems acquire during their natural growth. Whether it is also a meaningful statistical description depends on whether the collection of such irreversibly grown structures is amenable to a statistical mechanical study. In the case of causal and homogeneous networks [1], it has been shown that growing (non-equilibrium) networks can be assigned appropriate growth probabilities and studied within the framework of equilibrium statistical mechanics.

A trivial special case of a causal network is a Kinetic Growth Walk (KGW) [3-8] which may be considered as an ordered collection of 'nodes' (linear chain of monomers) such that two of them are of 'degree' one (end monomers) while the others are of 'degree' two each. It generates a Self Avoiding Walk (SAW) [2] configuration, step-by-step, by choosing one of the allowed steps at random with equal probability. Since the number of allowed steps at the growing end of the walk is a variable that depends on the configuration grown already, such a step-by-step growth is locally biassed. Averages taken over a collection of KGWs, uncorrected for the cumulative bias acquired during its growth, clearly establish the fact that KGW belongs to the same universality class as SAW [3-8]. In other words, like in the case of causal networks, the distinction between equilibrium (SAW) and non-equilibrium (KGW) configurations does not affect the scaling behaviour of the walk. This prompts us to take a closer view of how a collection of KGW configurations might be related to an (equilibrium) ensemble of SAWs.

In fact, the bias-uncorrected KGW on a honeycomb lattice has been shown [9] to be exactly equivalent to a canonical ensemble of Interacting SAWs (ISAWs) [10] at the bath temperature proportional to $(ln2)^{-1}$. These facts, put together, could be misconstrued to imply a general equivalence of KGW with ISAW at some bath temperature. Except for a hint [11] that kinetic walks may not be equivalent to canonical walks, no definitive study of this equivalence relation is available in the literature. In this paper, we present Monte Carlo evidence to show that such an equivalence does not exist on a square as well as on a triangular lattice. This is an explicit demonstration of the fact that the KGW is not, in general, equivalent to ISAW at any bath temperature.

\subsection{Definitions}

Starting from some arbitrarily chosen site on a regular lattice of coordination number $z$, there are $z$ possibilities for making the first step. For all subsequent non-reversal steps, we have a maximum of $z-1$ possibilities to choose from. If, at any stage, we choose a step that takes us to a site already visited, we discard the process and start afresh from the beginning. This ensures that every site in the walk is visited once, and once only. It is clear that the probability of generating a successful $N$-step walk configuration  can be written as 
\begin{equation}
P(N) = \frac{1}{z}\left( \prod _{k=2}^{N}\frac{1}{z_k}\right)
\end{equation}
where $z_k$ ($= 1, 2, 3,\cdots ,z-1$) is the number possibilities available for the $k^{th}$ step. If $z_k$ is taken to be $z-1$ for all the steps, then all the walk configurations are gernerated with equal probability, say $P_{SAW}(N)$, and are said to form an ensemble of Self Avoiding Walks (SAWs) [2]. 

On the other hand, if $z_k$ is the actual number of possibilities available for the $k^{th}$ step, then different configurations of the walk, called Kinetic Growth Walk (KGW) [3-8], are generated with different probabilities.
\begin{eqnarray}
P_{KGW}(N) & = & \left( \prod _{k=2}^{N}\frac{z-1}{z_k}\right)\left[\frac{1}{z}\left( \prod _{k=2}^{N}\frac{1}
                 {z-1}\right) \right] \\ 
           & = & \exp \left(\sum _{k=2}^{N}ln\left[\frac{z-1}{z_k}\right]\right) P_{SAW}(N)
\end{eqnarray}
Whenever $z_k < z-1$, we encounter non-bonded nearest neighbors, called {\it contacts}; the number of such {\it contacts} encountered at the $k^{th}$ step is given by $m_k = (z-1)-z_k$, which when added up give the total number of contacts, $n$, in the walk. 

\subsection{KGW on a honey-comb lattice}

The probability of generating an $N$-step KGW configuration with $n$ contacts is given by [9],
\begin{equation}
P_{KGW}(N) = e^{nln2}P_{SAW}(N)
\end{equation}
which follows from the fact that the walk has no choice but to take the only available step after making a contact. Since the total number of $N$-step SAWs, $\aleph _{SAW}(N)$, can be written as
\begin{equation}
\aleph _{SAW}(N) = \sum _{n=0}^N g(n)
\end{equation}
where $g(n)$ is the number of SAWs with $n$ {\it contacts}, the total number of KGWs on a honey-comb lattice is given by
\begin{equation}
\aleph _{KGWhc}(N) = \sum _{n=0}^N g(n)e^{-n(-\epsilon)ln2}
\end{equation}
Thus, the KGW corresponds to the canonical ensemble of ISAWs [10] realized at an inverse temperature, $\beta _{hc} = ln2$ if the magnitude of the energy per contact, $\epsilon$, is taken to be unity. At any other 'bath' temperature, $\beta$, KGW is identical to an ISAW at a lower temperature , $(\beta + \beta _{hc})^{-1}$. Algorithmically, we {\it first} choose an available site at random (KGW rule) and {\it then} assign the Boltzmann factor $e^{\beta}$ if a contact was made. 

\subsection{KGW on a square lattice}

The number of sites available for the current step is either one or two or three depending on whether, in the previous step, the walk had made a double contact or a single contact or no contact respectively. The probability of generating an $N$-step KGW configuration that has made $n_1$ single contacts and $n_2$ double contacts during its growth is therefore given by
\begin{equation}
P_{KGW}(N) = e^{n_1ln(3/2)+n_2ln(3)}P_{SAW}(N)
\end{equation}
Note that contacts made by the $N^{th}$ step, if any, is not accounted for by the growth probability $P_{KGW}(N)$. So, the total number of contacts made by the walk configuration is 
\begin{equation}
n = \left\{ \begin{array}{r@{\quad: \quad}c}
            n_1 + 2n_2 & \mbox{no contact at last step}\\ 
           (n_1+1)+ 2n_2 & \mbox{single contact at last step}\\
            n_1 + 2(n_2+1) & \mbox{double contact at last step}
            \end{array}\right.
\end{equation}            
It is not necessary that all the configurations with the same total number of contacts will be grown with the same probability because, growth being a history-dependent process, they could realize different pairs of values $\{ n_1,n_2\}$ for the same value of $n$. 

For example, consider an $8$-step configuration encoded as $00323212$ [12] that has a total of three contacts ($n_1 = 1; n_2 = 1$). The same configuration when grown in the reverse direction, encoded $03010122$, has only encountered single contacts ($n_1 = 3; n_2 = 0$). 

The fact that the total number of SAW configurations with $n$ contacts, $g(n)$, may be written as a double sum,
\begin{equation}
g(n) \equiv \sum _{n_1,n_2} \delta ([n_1+2n_2]-n),
\end{equation}
suggests that the total number of KGW configurations with $n$ contacts, $h(n)$, may also be given a similar expression - namely,
\begin{equation}
h(n) \equiv \sum _{n_1,n_2} e^{-[n_1(-\epsilon)\beta _1 + n_2(-2\epsilon)\beta _2]}\delta ([n_1+2n_2]-n)
\end{equation}
where $\beta _1 \equiv ln(3/2)$ and $\beta _2 \equiv (1/2)ln(3)$. They may be interpreted as the inverse 'temperatures' associated with the single and the double contacts respectively if $\epsilon \ (=1)$ is taken to be the (additive) energy per contact. Since ($\beta _2 > \beta _1$), a double contact is 'colder' than a single contact!

Configurations that are identical but grown in opposite directions are also counted into the expression for $h(n)$ by virtue of the fact that they correspond to different pairs $\{n_1,n_2\}$. It is therefore clear that we have a distribution of $\{n_1,n_2\}$'s (or equivalently, a distribution of dimensionless contact-energy, ${\cal E}(n_1,n_2) \equiv n_1\beta _1 + 2n_2\beta _2$ arising due to the history-dependence of the growth process. 

A canonical relation between $h(n)$ and $g(n)$ can be defined only if an average inverse 'temperature' $\beta (n)$ for a configuration with $n$ contacts can be defined:
\begin{equation}
h(n) = e^{-n(-\epsilon)\beta (n)}g(n) \quad \mbox{where} \quad \beta (n) \equiv \frac{<n_1>}{n}\beta _1 + \frac{2<n_2>}{n}\beta _2
\end{equation}
where $<n_1>$ and $<n_2>$ denote the average number of single and double contacts respectively for a configuration with $n$ contacts and may, in general, depend on $n$. Averaging out the history-dependence this way leads to the following expression for the total number of KGW configurations on a square lattice:
\begin{equation}
\aleph _{KGWsq}(N) = \sum _{n=0}^N g(n)e^{-n(-\epsilon)\beta (n)}
\end{equation}
It is still not in the canonical form, in contrast to that for a honey-comb lattice (Eq.(6)),  because $\beta (n)$ is an intrinsic geometric parameter that characterizes a configuration with a given number of contacts. and not the inverse canonical (or 'bath') temperature. 

%
%
\begin{figure}
\includegraphics[width=3.5in,height=2.75in]{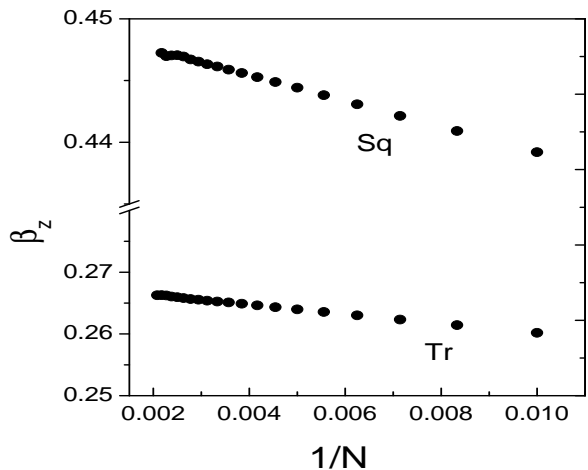}
\caption{Size dependence of the average $\beta _{z}$ for KGW. The asymptotic value of $\beta _{z}$ is of the order of $0.448$ for a square lattice (Sq), and is of the order of $0.267$ for a triangular lattice (Tr). }
\end{figure}
%
%
%
%
\begin{figure}
\includegraphics[width=3.5in,height=2.75in]{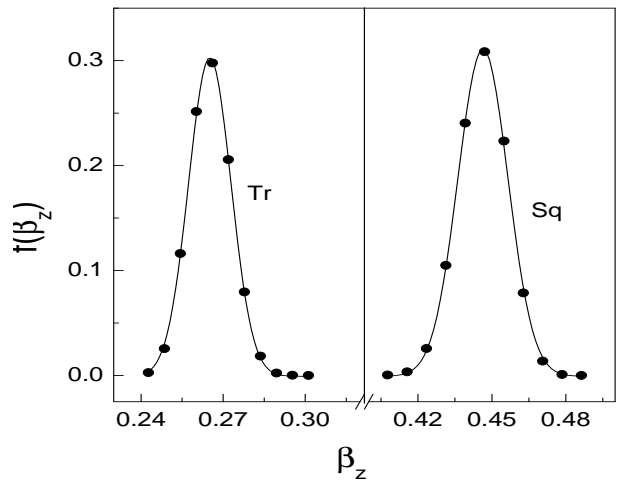}
\caption{The distribution of $\beta _z$ values for $300$-step KGW configurations on a triangular lattice as well as on a square lattice. It is a truncated distribution to which a Gaussian could be fitted. The relevant parameters are (i) peak position $\sim 0.265 (Tr); 0.446 (Sq)$; (ii) width $\sim 0.0155 (Tr); 0.0203(Sq)$.}
\end{figure}
%
%
%
%
\begin{figure}
\includegraphics[width=3.5in,height=2.75in]{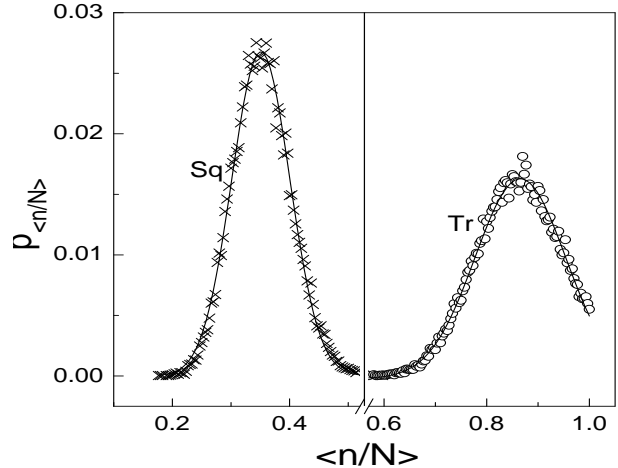}
\caption{The probability distribution for the number of contacts per step, $n/N$, of a KGW for $N=300$. A gaussian could be fitted to this distribution for $300$-step KGWs on a square lattice (Sq) as well as on a triangular lattice (Tr). The relevant parameters are, (i) peak value $\sim 0.351(Sq); 0.866 (Tr)$; (ii) width $\sim 0.099 (Sq); 0.1766 (Tr)$}
\end{figure}

A further averaging of the $\beta (n)$ values leads to the canonical expression
\begin{equation}
\aleph _{KGWsq}(N) = \sum _{n=0}^N g(n)e^{-n(-\epsilon)\beta _{sq}}
\end{equation}
Identifying $\beta _{sq}$ as the inverse 'bath' temperature is, however, not justifiable because a canonical sampling of SAWs with $\beta_{sq}$ may not lead to KGWs. Such an interpretation is unique to honey-comb lattice only. 

The Monte Carlo estimates of $\beta _{sq}$, plotted in Fig.1 for various walk-lengths, $N$, extrapolate to the asymptotic value $\sim 0.448$. The distribution of $\beta _{sq}$ values for a given walk-length, $N = 300$ in Fig.2, is {\it a priori} expected to be a truncated one because $\beta _{sq}$ is bounded between $\beta _1$ and $\beta _2$. However, a Gaussian (width $\sim 0.0203$, peak position $\sim 0.446$) could be fitted to the distribution.

The question remains whether $\beta _1^{-1}$ or $\beta _2^{-1}$ could be identified as the 'bath' temperature. The fact that KGW belongs to the same universality class as SAW does not provide a reason to a choose between the two, though they are both above the expected $\theta$-temperature ($\simeq 1.54$). Moreover, such an arbitrary choice also leads to non-additive contact energy.  

\subsection{KGW on a Triangular lattice}

A KGW on a triangular lattice is, in general, characterized by four inverse 'temperatures' ($\beta _1 = ln(5/4), \beta _2 = (1/2)ln(5/3), \beta _3 = (1/3)ln(5/2)$ and $\beta _4 = (1/4)ln(5)$) associated respectively with a single, double, triple and quadruple contacts. It is interesting to note that $\beta _4$ is greater than the expected $\theta$-point value ($\sim 0.375$), whereas the others are less.

The total number of KGWs on a triangular lattice may be given an approximate canonical form, similar to the one for KGWs on a square lattice (Eq.13):
\begin{equation}
\aleph _{KGWtr}(N) = \sum _{n=0}^N g(n)e^{-n(-\epsilon)\beta _{tr}}
\end{equation}
where the asymptotic value of $\beta _{tr}$ ($\sim 0.267$) could be obtained by extrapolating the data presented in Fig.1. The truncated distribution of $\beta _{tr}$ values, plotted in Fig.2 for walk-length $N = 300$, is again fitted to a Gaussian, which peaks at $\sim 0.265$ and whose width is $\sim 0.0155$.

\subsection{Summary}

In general, a KGW on a two dimensional lattice of coordination number $z$ is characterized by a set of $(z-2)$ inverse 'temperatures', $\beta _m$ ($m = 1, 2,\cdots ,z-2$), defined by
\begin{equation}
\beta _m \equiv \frac{1}{m}ln \left( \frac{z-1}{z-[m+1]}\right) \quad ; \quad m = 1, 2,\cdots , z-2
\end{equation}
whose weighted average gives the canonical value $\beta _z$:
\begin{equation}
\beta _z = \sum _{n=1}^N \sum _{m=1}^{z-2} \left( \frac{<n_m>}{n}\right) \beta _m
\end{equation}
where $<n_m>$ denotes the average number of $m$-contacts made. Note that $\beta _m = 0$ for walks without contacts ($m=0$). The dimensionality dependence comes through the realizable values of $n_m$s. As pointed out earlier, it is more appropriate to interpret $\beta _z$ as an intrinsic geometric parameter that characterizes the KGW than to interpret it as an inverse 'bath' temperature. However, it may serve as a simple scale-shift parameter for the inverse 'bath' temperature if all the KGW configurations are reweighted so as to correspond to $\beta _z$. The reason why KGW turns out to be equivalent to SAW could be that the entire truncated distribution of $\beta _z$s lies below the corresponding $\theta$-point value.

One of the authors (M.P) acknowledges the research grant from Council of Scientific and Industrial Research (CSIR No:9/532(19)/2003-EMR-I) India.

$^*$ Author for correspondence ({\it e-mail}: slnoo@magnum.barc.ernet.in).

\begin{enumerate}

\item P. Bialas, Z. Burda, J. Jurkiewicz, A. Krzywicki, {\it Phys. Rev.} E{\bf 67}, 66106 (2003);
      P. Bialas, preprint arXiv:cond-mat/0403669 (2004), and references therein.
\item P. G. deGennes, {\it Scaling Concepts in Polymer Physics} (Cornell University Press, Ithaca, 1979);
      C. Vanderzande, {\it Lattice Models of Polymers} (Cambridge University Press, 1998)
\item I. Majid, N. Jan, A. Coniglio and H. E. Stanley, {\it Phys. Rev. Lett}, {\bf 52}, 1257 (1084)
\item S. Hemmer and P. C. Hemmer, {\it J. Chem. Phys.} {\bf 81}, 584 (1984); {\it Phys. Rev.}, A{\bf 34},
      3304 (1986) 
\item L. Pietronero, {\it Phys. Rev. Lett.} {\bf 55}, 2025 (1985)
\item J. W. Lyklema and K. Kremer, {\it Phys. Rev. Lett.} {\bf 55}, 2091 (1985)
\item L. Peliti, {\it J. Phys.(Paris) Lett.} {\bf 45} L925 (1984)
\item A. Coniglio, N. Jan, I. Majid and H. E. Stanley, {\it Phys. Rev.} B{\bf 35}, 3617 (1987)
\item P. H. Poole, A. Coniglio, N. Jan and H. E. Stanley, {\it Phys. Rev.} B{\bf 39}, 495 (1989)
\item A SAW configuration is an athermal object because it is not assigned an {\it energy} value.   
      On the other hand, if it is assigned an {\it energy} value, it may be considered as an object sampled at  
      infinite temperature. A SAW that is assigned an energy value and sampled at a given bath 
      temperature is called an Interacting Self Avoiding Walk (ISAW). See, H. Saleur, J. Stat. Phys. {\bf 45}, 
      419 (1986); B. Duplantier and H. Saleur, Phys. Rev. Lett. {\bf 59}, 539 (1987).  
\item A. L. Owczarek and T. Prellberg, Physica A{\bf 260}, 20 (1998)
\item The codes $0, 1, 2$ and $3$ stand for steps in the $+x, +y, -x$ and $-y$ directions respectively.

\end{enumerate}

\end{document}